\documentstyle[aps,12pt]{revtex}

\begin{document}
\author{E.N. Glass\thanks{%
Permanent address: Physics Department, University of Windsor, Ontario N9B
3P4, Canada} and J.P. Krisch}
\address{Department of Physics, University of Michigan, Ann Arbor, Michgan 48109}
\date{22 October 1999}
\title{Schwarzschild Atmospheric Processes:\\
A Classical Path to the Quantum\thanks{%
The original version of this paper received Honorable Mention as a 1999
Gravity Research Foundation Essay}}
\maketitle

\begin{abstract}
\\\\%
--------------------------------------------------------------------------------------------------%
\\We develop some classical descriptions for processes in the Schwarzschild
string atmosphere. These processes suggest relationships between macroscopic
and microscopic scales. The classical descriptions developed in this essay
highlight the fundamental quantum nature of the Schwarzschild atmospheric
processes.\\%
--------------------------------------------------------------------------------------------------%
\newpage\ 
\end{abstract}

\section{Atmosphere and Diffusion}

General Relativity provides a wealth of information about the events
surrounding us over a large range of scales in both space and time. We have
analytic solutions to Einstein's field equations which span distances from
the size of the Universe down to the size of a stellar black hole. When
contemplating quantum effects at classical boundaries such as the
Schwarzschild horizon, the scales become much smaller. The classical
electron, a point particle, may perhaps be resolved into Planck scale
strings. String bits have lengths proportional to the Planck length $\sqrt{%
\hbar G/c^3}$ = $1.6\times 10^{-35}$ $m$, and $10^{20}$ strings bits fit
into a classical electron radius. On this smaller scale, the classical field
equations can also provide insights in terms of correspondence limits. The
physics of a Schwarzschild object is an arena where both large and small
distance scales meet and ideas about macroscopic and microscopic
correspondence can be developed and applied. In this essay we will discuss
some classical and quantum aspects of the atmosphere around a Schwarzschild
object.

The Schwarzschild metric, characterized by a single mass parameter, has
played an important role in understanding general relativistic solutions.
Vaidya \cite{vaidya} showed that allowing the mass parameter to be a
function of retarded time created a null fluid atmosphere. Glass and Krisch 
\cite{ed-jean2} discovered that allowing the same parameter to be a function
of the radial coordinate creates a string fluid atmosphere. In a spherically
symmetric spacetime with metric $ds_{GK}^2=A(u,r)du^2+2dudr-r^2d\Omega ^2$,
the field equations relate the mass function $m(u,r)$ to the density of the
string fluid atmosphere by 
\begin{equation}
\partial _rm=4\pi r^2\rho .  \label{dens}
\end{equation}
If one assumes that 
\begin{equation}
\partial _um=4\pi D(r)r^2\partial _r\rho   \label{dr-dens}
\end{equation}
then the density obeys a diffusion equation 
\begin{equation}
\partial _u\rho =r^{-2}\partial _r[D(r)r^2\partial _r\rho ]
\label{rho-diffu}
\end{equation}
with variable diffusivity $D(r)$. (In real diffusion problems, variable
diffusivities are the rule rather than the exception \cite{gh-lang}.) A
string fluid is a continuum description of a collection of quantum string
bits. A link between the macroscopic and microscopic pictures is reflected
in the density transport rule. The density diffuses in the Glass-Krisch
atmosphere with $A(u,r)=1-2m(u,r)/r.$

Since $m$ describes the horizon position, the behavior of the horizon is
linked to the string fluid density. Assume that the mass is diffusing in a
spherically symmetric spacetime of spatial dimension $\delta $ and obeys a
diffusion equation 
\begin{equation}
\partial _um=r^{1-\delta }\partial _r[\tilde{D}(r)r^{\delta -1}\partial _rm].
\label{m-diffu}
\end{equation}

Using Eqs.(\ref{dens}) and (\ref{dr-dens}) and demanding consistency with
Eq.(\ref{rho-diffu}) we find the diffusivities are equal and must have the
form 
\begin{equation}
D(r)=\tilde{D}(r)=D_{_0}r^{-(1+\delta )}.  \label{d-func}
\end{equation}
If the diffusivity is constant 
\begin{equation}
\delta =-1,\ \ D(r)=D_{_0},  \label{const-dfunc}
\end{equation}
and the flowspace for mass transport then has dimension $\delta =-1$.
Negative dimensions have occurred in calculations of critical behavior on
random surfaces \cite{b-dup},\cite{kos-meh},\cite{ka-kos}. In this case, the
negative dimension describes an internal space embedded in an external
manifold of possibly different dimension. Another possible explanation of
dimension $\delta $ follows from Visser's idea \cite{visser} of an
hydrodynamic metric. He showed that the equation of fluid flow in a flat
spacetime could be formally written in terms of a metric related to the
hydrodynamic parameters of the flow. Visser's idea of an hydrodynamic
flowspace is very similar to the idea of an internal space. All that the
diffusion equation for $m(u,r)$ says about the flowspace is that $\sqrt{-g}%
=r^{\delta -1}f(\vartheta )$. The flowspace metric could be in the standard
spherical form with $\delta =-1$, but it could also have any form consistent
with Eq.(\ref{m-diffu}).

The mass solution $m=4\pi c(u)r^\alpha $ provides an interesting example of
dimensional relations. Substituting into both the mass and density diffusion
equations and demanding consistency, we find that the dimension of the mass
flowspace is the negative of the dimension of the density space and that the
diffusivity is determined: 
\[
\delta =-3,\ \ D(r)=D_{_0}r^2. 
\]

Physically the mass diffusion can be understood in terms of the position of
the event horizon . An horizon whose mass depends on $r$ creates a string
fluid atmosphere. If the mass begin to change, the horizon will also begin
to change its position in time. If it moves inward, then the string fluid
density, also a function of position, begins to diffuse to smaller $r$,
feeding string fluid into the regions close to the horizon. If the horizon
is growing then the string fluid diffuses outward to larger $r$.

\section{A Classical Analogy}

The string fluid is serving as a dissipative atmosphere around a
Schwarzschild horizon. The central Schwarzschild object draws upon the
string fluid, trying to maintain the horizon position. This is somewhat
analogous to the behavior of an inductive circuit with the string fluid
attempting to counter the changing horizon position as an inductor tries to
counter changes in magnetic flux. This analogy can be carried further by
noting that in a lossy transmission line, the current and voltage obey the
equations 
\begin{eqnarray}
LI,_t+RI &=&-V,_w  \label{trans_line} \\
CV,_t+GV &=&-I,_w\ ,  \nonumber
\end{eqnarray}
where the circuit parameters, $L$, $R$, $G$, and $C$, are all per unit
length. These first order equations are equivalent to a second order
telegrapher's wave equation for $I$ or $V$.

If we require that the transmission line have $R=C=0$ then these equations
are formally identical to equations (\ref{dens}) and (\ref{dr-dens}) with $%
I\rightarrow m,$ $V\rightarrow \rho $, $t\rightarrow u$, $w=1/r$, and where
the inductance and leakage inductance densities are given by 
\begin{eqnarray}
L &=&1/4\pi D(w)  \label{lc-analog} \\
G &=&4\pi /w^4.  \nonumber
\end{eqnarray}
A similar analogy with only resistance and capacitance has $G\rightarrow R$, 
$L\rightarrow C$, and $V\rightarrow m$. It is not surprising that the
atmosphere can be modeled as a transmission line since the string fluid and
null fluid each serve as a transmission medium for the other. An interesting
aspect of this classical analogy is the ''lossiness'' of the atmosphere,
either through resistance or leakage inductance.

\section{Diffusion and Dissipation}

The description of a string atmosphere diffusing inward as null radiation
moves outward classically models underlying quantum processes. The
dissipation in the electrical analogy models quantum friction. Quantum
dissipation is usually the result of coupling between a system and a complex
environment \cite{weiss}. The coupling causes information loss. There have
been several recent suggestions about the origin of quantum friction in
stringy systems. Ellis et al \cite{e-m-n} have suggested that when light
particles scatter from D-branes, neglecting the D-brane recoil results in
information loss. They also find, using Renormalization Group arguments,
that the D-brane wave function evolves diffusively. A related suggestion by
the same authors \cite{e-m-n2} considers the interaction of light particles
with an environment of spacetime foam. In this calculation, quantum friction
is due to couplings with unobserved massive string states. Both examples
explain quantum friction as due to information loss in a scattering process.
Diffusion is, of course, intrinsically dissipative since in a diffusive
process one may only predict and not retrieve. A possible origin for the
dissipation in the string atmosphere is through interactions between the
null radiation and the string fluid. This can be examined in a simple model
calculation. Consider a static model with $m=m(r)$ and a string fluid
density $\rho =\rho _0+k_1/r$. From (\ref{dens}) the string mass is $%
m(r)=m_0+(4\pi /3)r^3\rho _0+2\pi k_1r^2$. If the mass is allowed to be a
function of retarded time and $D(r)=D_{_0}$, this particular density will
not change but the mass is now \newpage\ $m(u,r)=m_0+(4\pi /3)r^3\rho
_0+2\pi k_1(r^2-2D_{_0}u)$. We can identify the first three terms as the
string mass and the last term can be interpreted as the net flux through an $%
r=constant$ spacelike 3-surface \cite{ed-jean2}. The net flux is due to the
outward moving radiation and the inward diffusing string bits. The constant
diffusivity parametrizes the net flux. A diffusivity is a measure of the
resistance offered to the diffusing medium by the surrounding environment 
\cite{morse-fesh} so this term could represent the interaction between the
null fluid created by $\partial _um$ and the string fluid. There are several
other possible sources of friction. The complete extension of the Vaidya
metric generates fluid with both radial and transverse stress. The
transverse stress can be attributed to dust \cite{ed-jean2} or to a non-zero
magnetic contribution to the string bivector \cite{letel}. String-dust
scattering could be a source of information loss. Another possible loss
mechanism is the snapping of tidally stretched string bits \cite{b-h-o}, the
snap and loss of potential energy, sending a disturbance through the
surrounding environment.

\section{Scale and Other Transport Processes}

We have seen in the previous section that the diffusion can provide
significant insights into the behavior of the horizon. We assumed a
diffusion equation for the atmospheric string density and found the
resulting diffusion equation for the mass; for $D(r)=D_{_0}$ each scale with
the Boltzmann scaling variable $\eta $ \cite{ghez}, where $\eta ^2=\frac{r^2%
}{4D_{_0}u}$. This scaling variable is traditionally associated with simple
diffusive mass transport. Another source of similarity behavior and scaling
variables in the Schwarzschild system is the metric itself \cite{ed-jean3}.
Because the atmospheric string fluid lives on a 2-dimensional world sheet in
the ($u,r$) plane, the similarity behavior of the matter surfaces is of
interest. Starting with an assumed scaling behavior, we try to develop an
associated mass transport rule.

The Glass-Krisch metric can be written in terms of unit vectors as 
\[
g_{ab}^{GK}=\hat{v}_a\hat{v}_b-\hat{r}_a\hat{r}_b-\hat{\vartheta}_a\hat{%
\vartheta}_b-\hat{\varphi}_a\hat{\varphi}_b 
\]
(see \cite{ed-jean2} for details). The ($u,r$) world sheet is spanned by
unit vectors $\hat{v}_a$ and $\hat{r}_a$ and is scaled by 
\begin{equation}
{\cal L}_\xi (\hat{v}_a\hat{v}_b-\hat{r}_a\hat{r}_b)=2\mu (\hat{v}_a\hat{v}%
_b-\hat{r}_a\hat{r}_b).  \label{vr-scale}
\end{equation}
Similarly, the orthogonal ($\vartheta ,\varphi $) 2-surfaces are spanned by $%
\hat{\vartheta}_a$ and $\hat{\varphi}_a$ and are scaled by 
\begin{equation}
{\cal L}_\xi (\hat{\vartheta}_a\hat{\vartheta}_b+\hat{\varphi}_a\hat{\varphi}%
_b)=2\nu (\hat{\vartheta}_a\hat{\vartheta}_b+\hat{\varphi}_a\hat{\varphi}_b).
\label{thetphi-scale}
\end{equation}
The similarity vector which preserves the distinct ($u,r$) 2-surfaces of the
matter distribution is 
\[
\xi ^a\partial _a=[\nu u_{_0}+(2\mu -\nu )u)]\partial _u+\nu r\partial _r. 
\]
The scale symmetry of Eqs.(\ref{vr-scale}) and (\ref{thetphi-scale})
requires a first order differential constraint on $g_{uu}=A=1-2m(u,r)/r$. 
\begin{equation}
(u_{_0}+\kappa u)A_{,u}+rA_{,r}=(1-\kappa )A.  \label{kin_constraint}
\end{equation}
where $\kappa :=2\mu /\nu -1$. The assumption of a similarity transform on
the matter 2-surfaces has imposed a mass transport rule. This is
recognizable as a first order form of the telegrapher's equation. A lossy
transmission line analogy is possible here (\cite{morse-fesh} p. 219). In
terms of coordinates $t=\ \kappa ^{-1}$ln$(u_{_0}+\kappa u)$ and $q=$ln$(r)$%
, the second order form of Eq.(\ref{kin_constraint}) is 
\[
A_{,tt}-A_{,qq}+(\kappa -1)(A_{,q}-A_{,t})=0 
\]
where the presence of $A_{,t}$ indicates damping of wave motion.

When $\kappa =1$ and $(\mu ,\nu )=(1,1)$ then A obeys a simple wave equation
on the flat tangents to the string 2-space. There is no dissipation.
Choosing $\kappa =1$ makes the map homothetic on the entire spacetime, $%
{\cal L}_\xi g_{ab}=2g_{ab}$. Since density and mass are related by Eq.(\ref
{dens}), we obtain a second order equation for the density from Eq.(\ref
{kin_constraint}) 
\[
\rho _{,tt}-\rho _{,qq}+(\kappa -1)\rho _{,t}=(\kappa +3)\rho _{,q}+2(\kappa
+1)\rho 
\]
and we again find the dissipative term vanishing for a homothetic map.

Other parameter values will include diffusive effects. For example $(\mu
,\nu )=(1/2,1)$, $\kappa =0$, preserves the scale of the string 2-surfaces
while acting homothetically on the orthogonal space. For these parameters
one can show from Eq.(\ref{kin_constraint}) that an associated second order
equation is (with a more well known form of the telegrapher's equation) 
\begin{equation}
A_{,uu}-3A_{,u}/u_{_0}-(r/u_{_0})^2\nabla ^2A=-2A/u_{_0}^2.  \label{teleg}
\end{equation}

\section{Conclusion}

To summarize, we have seen that the string fluid atmosphere around a
Schwarzschild black hole can have profound effects on the mass transport
which drives the horizon position. Common elements of the mass transport
considered are the existence of a scaling variable and analogies to
classical dissipative systems. This is an especially interesting result to
hold across several models since t'Hooft \cite{thooft} has suggested the
necessity for dissipation of information as an ingredient of a theory of
quantum gravity. The absence of dissipation in the mass transport associated
with a homothetic map is also very interesting, given his suggestion. The
presence of both the diffusion equation and the telegrapher's equation \cite
{kac} is also highly suggestive of the underlying quantum nature of the
atmospheric processes since both of these equations have been used to link
macroscopic and microscopic physics. The classical descriptions developed in
this essay highlight the fundamental quantum nature of the Schwarzschild
atmosphere.

\end{document}